\documentclass[twocolumn]{aastex61}
\submitjournal{ApJ}
\shorttitle{Simulating the Microlensed X-Ray Emission from Quasars}
\shortauthors{Krawczynski, Chartas, Kislat}
\usepackage{amsmath}

\def\rxj{{RX J1131$-$1231}}

\usepackage{epstopdf}
\begin{document}


\title{The Effect of Microlensing On the Observed X-ray Energy Spectra of Gravitationally Lensed Quasars}

\correspondingauthor{Henric Krawczynski}
\email{krawcz@wustl.edu}
\author{H. Krawczynski}
\affil{Physics Department and McDonnell Center for the Space Sciences, Washington University in St. Louis, 1 Brookings Drive, CB 1105, St. Louis, MO 63130}
\author{G. Chartas}
\affil{Department of Physics and Astronomy, College of Charleston, Charleston, SC, 29424, USA}
\affil{Department of Physics and Astronomy, University of South Carolina, Columbia, SC, 29208}
\author{F. Kislat}
\affil{Department of Physics \& Space Science Center University of New Hampshire, 
Durham, NH 03824}
\begin{abstract}
The {\it Chandra} observations of several gravitationally lensed quasars show evidence 
for flux and spectral variability of the X-ray emission that is uncorrelated between images and is 
thought to result from the microlensing by stars in the lensing galaxy. 
We report here on the most detailed modeling of such systems 
to date, including simulations of the emission of the Fe K$\alpha$ fluorescent radiation 
from the accretion disk with a general relativistic ray tracing code, the use of realistic 
microlensing magnification maps derived from inverse ray shooting calculations, 
and the simulation of the line detection biases. We use lensing and black hole parameters 
appropriate for the quadruply lensed quasar RX~J1131$-$1231 ($z_s\, =$~0.658, $z_l\, =$~0.295), 
and compare the simulated results with the observational results. The simulations cannot fully 
reproduce the distribution of the detected line energies indicating that some of the 
assumptions underlying the simulations are not correct, or that the simulations are 
missing some important physics. We conclude by discussing several possible explanations.  
\end{abstract}
\section{Introduction \label{sec:intro}}
The gravitational lensing of the emission from distant quasars produces several macroimages of the quasar
with possible substructure on micro-arcsecond and milli-arcsecond scales owing to the discrete nature of the stellar
component and the clumpy nature of the dark matter components of the lensing galaxy, respectively 
\citep[see the review by][]{Wamb:06}. The micro, mili, and macro images correspond to the stationary 
paths of the Fermat potential (or time delay function) generated by the stars, clumpy dark matter and stars, 
and the entire galaxy, respectively \citep[e.g.][]{Pacz:86,Blan:86,Wamb:92,Yone:03}.
The motion of the deflectors relative to the line of sight results in uncorrelated brightness fluctuations 
of the macroimages as the caustic folds and cusps created by the granular mass distribution move 
across the source, each flare corresponding to the appearance or disappearance of two microimages.

Observations of the microlensing brightness fluctuations of quasars have been used to 
constrain the sizes of the quasar emission regions as the amplitude of the microlensing 
flux variability depends on the ratio of the angular radius of the source to the Einstein 
radius $\alpha_0$ (two angles). The latter is given by: 
\citep[e.g.][]{1992grle.book.....S}:
\begin{equation}
\alpha_0\,=\,\sqrt{\frac{4\,G\,<\!\!M_*\!\!>}{c^2} \frac{D_{\rm LS}}{D_{\rm L}D_{\rm S}}}.
\end{equation}
where $<\!\!M_*\!\!>$ is the average mass of the deflecting stars, and $D_{\rm LS}$, $D_{\rm L}$, and
$D_{\rm S}$ are the angular diameter distances between the lens and the source, 
the observer and the lens, and the observer and the source, respectively.    
Whereas the observations indicate that the optical/UV bright portions of the accretion disks might 
be significantly larger than predicted by thin disk theory \citep{2010ApJ...712.1129M}, 
the X-ray bright accretion disk coronae seem to be extremely compact, i.e.\ 
smaller than 30~$r_{\rm g}$ with $r_{\rm g}=GM/c^2$ being the gravitational 
radius of the black hole \citep{Char:09,Dai:10,Morg:12,Mosq:13,Blac:15,MacL:15}.
These X-ray microlensing constraints are independent and complimentary to
spectral \citep[e.g.][]{Wilm:01,Fabi:09,Zogh:10,Park:14,Chia:15} 
and reverberation \citep[e.g.][]{Zogh:14,Cack:14,Uttl:14} constraints 
on the corona sizes of nearby (unlensed) Narrow Line Seyfert I (NRLS I) active galactic nuclei (AGNs).  

Even well before the observational discovery of microlensing flux variability, various authors remarked on 
the possibility of using quasar microlensing to constrain the stellar component \citep[e.g.][]{Webs:91,Lewi:96}. 
Several years later, \citet{Sche:02} used the inverse ray shooting (RS) method \citep{Youn:81,Pacz:86,Schn:87,KRS:86,Wamb:99} 
to generate a series of microlensing magnification maps for different ratios of stellar to dark matter masses. 
The authors found that the smooth mass component enhanced the amplitude of the brightness fluctuations.
Comparing the flux variations of the images corresponding to the saddle points of the
Fermat potential to those of the images corresponding to the minima of the Fermat potential
showed that saddle point images exhibited larger brightness fluctuations than minimum images
in agreement with the earlier result of \citet{Witt:95,Metc:01}. 

More recently, \citet{Pool:12} analyzed 61 observations of 14 quadruply lensed quasars observed 
with the  {\it Chandra} X-ray satellite with the aim to constrain the stellar to dark matter mass ratio.
They argued that X-ray observations were better suited than optical observations 
\citep[e.g.][]{Koch:04,Sche:04b} as the angular extents of the X-ray emission regions were 
smaller than the Einstein radius of the stars. 
The authors used for each image the convergence $\kappa$ and shear $\gamma$ from the analysis 
of the positions of the macroimages, and ran a series of RS simulations for different stellar convergence 
to total convergence fractions. Using the magnification probability distributions from the RS simulations
and the apriori probability distribution of the intrinsic quasar brightness fluctuations from deep 
studies of the cosmic X-ray background, they inferred a mean stellar (dark matter) contribution 
to the total convergence of 7\% (93\%) 
\citep[see][for related studies]{Sche:14,Jime:15}.

In this paper, we focus on using the RS method to simulate the spectral shapes of the microlensed 
Fe~K$\alpha$ emission from quasars. The Fe~K$\alpha$ emission is thought to originate from a hot 
X-ray bright corona illuminating the accretion disk with high energy X-rays, prompting the emission 
of fluorescent Fe~K$\alpha$ photons \citep[see][and references therein]{2014SSRv..183..277R}. 
As the energies of the escaping Fe~K$\alpha$ photons depend on the  
Doppler and gravitational frequency shifts between emission and detection, the 
photon energies encode information about where the photons originated in the accretion disk 
and about the background spacetime.
A  caustic fold crossing the accretion disk selectively amplifies the emission from a narrow 
slice of the accretion disk and produces energy spectra carrying the imprint of the 
Doppler and gravitational frequency shifts characteristic for the emission from this slice.
\citet{2002ApJ...568..509C} present simulated microlensed Fe~K$\alpha$ energy spectra accounting 
for Doppler and gravitational frequency shifts using the general parameterization of the magnification 
close to caustic folds \citep[e.g.][]{1992grle.book.....S}:
\begin{equation}
\mu/\mu_0\,=\,1+\frac{K}{\sqrt{y_{\perp}}}H(y_{\perp})
\label{para}
\end{equation}
with $y_{\perp}$ being (up to the sign) the distance from the fold located at $y_{\perp}\,=\,0$, 
$K$ encoding the lens properties, 
and the Heaviside step function $H(y_{\perp})\,=\,0$ for $y_{\perp}<0$ and $H(y_{\perp})\,=\,1$  
for $y_{\perp}\ge 0$.
The simulations predict distorted energy spectra with one or two spectral peaks.

The  {\it Chandra} observations of the quadruply lensed quasars \rxj\ 
($z_s\, =$~0.658, $z_l\, =$~0.295)  and SDSS~1004+4112 ($z_s\,=$~1.734, $z_l\,=$ 0.68), 
the double lens quasar QJ~0158$-$4325 ($z_s\,=$~1.294, $z_l\,=$~0.317) \citep{Char:12,Char:17},
and to lesser degree the quasar MG J0414+0534 ($z_s\,=$~2.64, $z_l\,=$~0.96) \citep{2002ApJ...568..509C}
seem to confirm these predictions, showing energy spectra deviating from an absorbed
power law model which can be fit with a model assuming one or two emission lines 
with line centroids changing from observation to observation.
Based on estimates of the masses of the three black holes and the effective velocity of the caustic patterns
across the quasars, \citet{Char:17} estimate that caustic folds cross the central 10$r_{\rm g}$ regions
of the accretion flows of the three quasars within 0.6 to 2.9 months, 
implying that some of the {\it Chandra} observations recorded energy spectra from different 
stages of the crossing of a single caustic fold moving across the inner portion of the accretion disk.  
The authors remark that the range of the observed line centroids and the ratio of the two line centroids
for energy spectra with two spectral lines can be used to constrain the inclination 
and spin of the black hole. 
\begin{table*}[tbh]
\centering
{\small 
\begin{center}
\begin{tabular}{|p{2.5cm}|p{1.7cm}|p{1.7cm}|p{1.7cm}|p{1.7cm}|}
\hline
Source & {\it Chandra} Pointings & Image &  
Spectra with detected lines ($>$99\% CL) & 
Spectra with 2 detected lines (both $>$99\% CL)\\ \hline
{RX J1131$-$1231}& 
{38}
& A (HS)     & 7 & 1 \\
&& B (HM)  & 5 & 0 \\ 
&& C (LM)  & 4 & 2 \\
&& D (LS)  & 2 & 0 \\ \hline 
%
{SDSS 1004+4112}& 
{10}
& A (HS)    & 1 & 1 \\
&& B (HM) & 1 & 0  \\
&& C (LM) & 1 & 0 \\
&& D (LS) & 2 & 0 \\ \hline 
%
{Q J0158$-$4325}& 
{12}
& A (HM)    & 2 & 1 \\
&& B (LM)  & 0 & 0 \\ \hline 
\end{tabular}
\end{center}}
\caption{Statistics of the {\it Chandra} Fe~K$\alpha$ detections. See \citep{Char:17} for details.}
\label{statistics}
\end{table*}

Various authors discuss more detailed simulations of the Fe~K$\alpha$ emission. 
\citet{Popo:03,Popo:06} use a general relativistic (GR) ray tracing code to predict the line shapes. 
The authors assume that the Fe~K$\alpha$ emissivity of the equatorial accretion disk 
either follows a power law dependence in the radial Boyer Lindquist coordinate $r$ or is 
proportional to the thermal emissivity. 
\citet{Jova:09} perform similar simulations
focussing on the impact of an absorber covering parts of the accretion disk. 
The authors use generic parameterizations of the magnification close to a caustic fold, 
and a few realizations of RS magnification maps. \citet{Nero:16} perform similar simulations 
and study the time evolution of the energy spectra during individual caustic crossings and 
posit that dense spectroscopic observations of such caustic crossings may present a 
new method for testing general relativity based on X-ray observations 
\citep[however, see][for a critical discussion of the magnitude 
and impact of astrophysical uncertainties]{Kraw:18}.  
In an earlier paper \citep[][called Paper I in the following]{Kraw:17} we use a GR raytracing code 
to simulate the illumination of a geometrically thin accretion disk by a lamppost corona
located at a height $h$ above the accretion disk. 
Combining these simulations with the magnification from Equation (\ref{para}), 
we discuss the distribution of the centroids and widths of the observed emission lines,
and in the case of energy spectra with two emission lines, the ratio of the two peak energies as 
function of the black hole spin, the inclination of the accretion disk relative to the observer,
the lamppost height $h$, the magnification parameter $K$, and the location and orientation
of the caustic folds. 

The present paper examines the statistical properties of the distorted Fe~K$\alpha$ lines,
i.e.\ the likelihood of detecting shifted or double spectral peaks.
We do so by combining the results of the GR raytracing simulations of the Fe~K$\alpha$ emission
with magnification maps from RS calculations. 
The study presented here is closely related to the earlier studies of 
\citet{Lewi:96,Sche:02,Sche:04,Pool:12,Sche:14,Jime:15} mentioned above as the observed
properties depend on the source parameters {\it and} the microlensing parameters. 
Eventually, we would like to examine the outer product of the parameter spaces describing the 
black hole, the accretion disk, the X-ray bright corona, and the microlensing, 
and evaluate which parameter combination describes the X-ray data best.
The comparison of the simulated and observed data would then constrain the properties of the
quasar and the lensing galaxy, including the properties of the stellar and dark matter components.  

\begin{figure*}[t]
\includegraphics[width=0.82\textwidth]{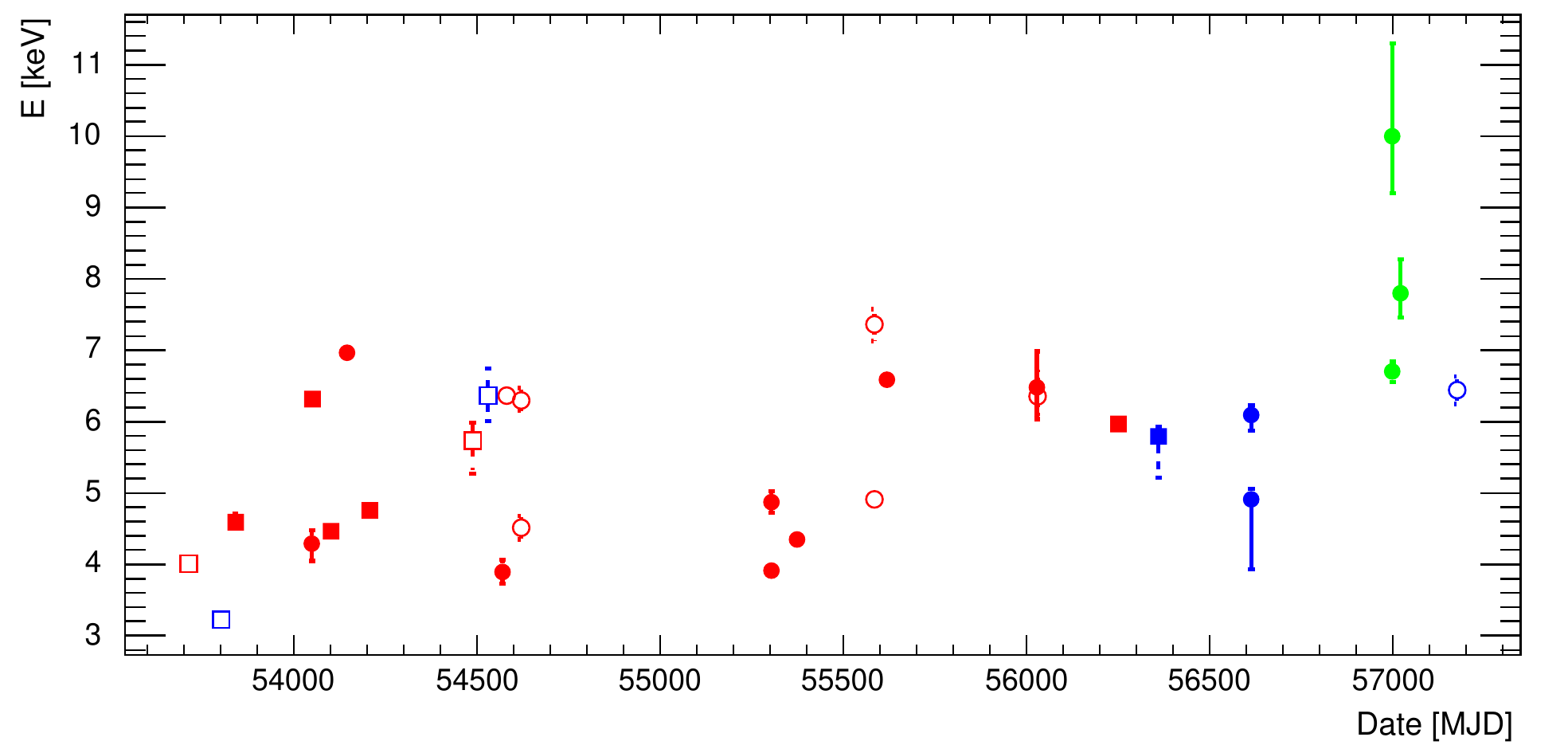}
\caption{\label{timeDep} Distribution of the detected rest frame line centroid energies as a function in time
for RX J1131$-$1231 (red), SDSS~1004+4112 (blue) and QJ~0158$-$4325 (green) for 
image A (full circles), image B (full squares), and if present, image C (open circles) and image D (open squares)
of the respective sources. We show only lines detected on a $>$99\% confidence level 
and 90\% confidence interval error bars.
The times of the images have been corrected for the measured or estimated time offsets 
of the individual images. RX J1131$-$1231:  Images B and C lead image A by 0.7 and 1.1 days, respectively, and 
image D lags image A by 91 days \citep{Tewes:13};
SDSS~1004+4112: Images B and C lead image A by 41 and 822 days, respectively, and image D lags image A by
1789 days \citep{Fohl:08,Fohl:13};
QJ~0158$-$4325: Image B lags image A by 14.5 days \citep{Faure:09}.
}
\end{figure*}
We limit the scope of this paper to considering only a small portion of the quasar and lensing parameter spaces,
and studying qualitatively if the simulated Fe~K$\alpha$ energy spectra can explain the observed
phenomenology.  
The observational results will be summarized in \S\ref{obs} and
the numerical simulation methods will be presented in \S\ref{methods}. 
Simulated magnification maps are presented in \S\ref{maps} and simulated Fe~K$\alpha$ 
energy spectra are discussed in \S\ref{fka}.  Finally, we use the simulated Fe~K$\alpha$ 
energy spectra to generate simulated {\it Chandra} data sets. We fit the simulated and 
observed {\it Chandra} spectra of \rxj\ in the same manner using an automated script 
described in \S\ref{chandra}.  The paper concludes with a summary and discussion 
of the results in \S\ref{discussion}.

In the following, we assume the cosmological parameters from the 2015 Planck release \cite{Plan:15}, 
i.e.\ a Hubble constant $H_0\,=$~67.8 km s$^{-1}$ Mpc$^{-1}$,  a matter density of 
$\Omega_{\rm m}/\Omega_{\rm c}\,=$~30.9\%, and a dark energy density 
$\Omega_{\Lambda}/\Omega_{\rm c}\,=$~ 69.1\%, with $\Omega_{\rm c}$ being the critical density.

If not mentioned otherwise, we use geometric units ($G$=$c$=1).
Distances in physical units are given in units of the gravitational radius $r_{\rm g}=G M_{\rm BH}/c^2$ 
with $M_{\rm BH}$ being the black hole mass. Denoting the angular momentum by $J$, the spin parameter 
is given by $a=J/c r_{\rm g} M_{\rm BH}=c J/ G M_{\rm BH}^2$. 
In units of $M_{\rm BH}$, the spin parameter $a$ can range from -1 to +1.
\section{Observational Results \label{obs}}
The three quasars with the best evidence for time variable Fe~K$\alpha$ lines 
are the quadruply lensed quasars \rxj\ and SDSS~1004+4112 
and the double lens quasar QJ~0158$-$4325 \citep[see][for details]{Char:17}.  
Table \ref{statistics} summarizes the number of {\it Chandra} observations, 
and for each macro image the number of observations with highly significant 
($>$99\% confidence level) single and double Fe~K$\alpha$ line detections.
We identify the images here with their letters A-D. For each image we note
if it corresponds to the higher magnification saddle point (HS),
the lower magnification saddle point (LS), the higher magnification minimum (HM), or
the lower magnification minimum (LM) of the Fermat potential.
The highly significant lines were detected in 12\%, 13\%, and 8\% of the energy spectra 
of  RX J1131$-$1231, SDSS~1004+4112, and QJ~0158$-$4325, respectively, 
Highly significant double peaks were detected in 2\%, 2.5\%, and 4\% of the energy spectra of the three sources, respectively.

Figure \ref{timeDep} shows for all three sources the rest frame centroids of the detected lines as a function of time.
Thirteen (43\%) of the detected lines have centroid energies below 5 keV, fifteen (50\%) have energies between 
5.6 keV and 7.4~keV, and two (7\%) have energies exceeding 7.6~keV.
Figure \ref{centroids} shows the distribution of the line centroids for image A of \rxj, exhibiting two peaks, one
at $\sim$4~keV, and one at $\sim$7~keV. 
Considering only the sample of $<5$~keV line detections, Figure \ref{timeDep} reveals some weak evidence for
a clustering of the detections in time, i.e.\ for image B of RX J1131$-$1231 three $<5$~keV lines were 
detected before MJD 54,250, and none afterwards, and  for image A of the same source only one  $<5$~keV 
line was detected before MJD 54,250, and four afterwards.
\section{Numerical Methods \label{methods}}
\subsection{General Relativistic Ray Tracing Code}
\label{rt}
Our GR ray tracing code has been 
described in \citep{2012ApJ...754..133K,2016PhRvD..93d4020H,Behe:2016,Behe:2017,Kraw:17}. 
It uses the Kerr metric in Boyer Lindquist (BL) coordinates $x^{\mu}=(ct,r,\theta,\phi)$ with $ct$ and $r$ 
in units of $r_{\rm g}$. A lamppost corona \citep[][]{1991A&A...247...25M} close to the spin axis of the 
black hole hovers at $r=5$~$r_{\rm g}$ above the black hole and emits photon packets isotropically in its rest frame. 
The photon packets are tracked until they impinge on a geometrically thin accretion disk 
extending from the Innermorst Stable Circular Orbit (ISCO) to 100 $r_{\rm g}$. 
The 0-component of the photon packet's wave vector $k^{\mu}$ is initially set to 1 and is used to keep track 
of the frequency shift along the packet's trajectory. Each photon packet represents 
a power law distribution of initially unpolarized corona photons with differential spectral 
index $\Gamma=1.7$ (from $dN/dE\propto E^{-\Gamma}$). 
The position and wave vectors are evolved forward in time by integrating the geodesic equation with an
adaptive stepsize Cash-Karp method. The code keeps track of the photon packets' polarization by storing 
the polarization fraction and vector. The polarization fraction is modified every time the photon
scatters, and the polarization vector is parallel transported along the photon packets' geodesic.
\begin{figure}[t]
\plotone{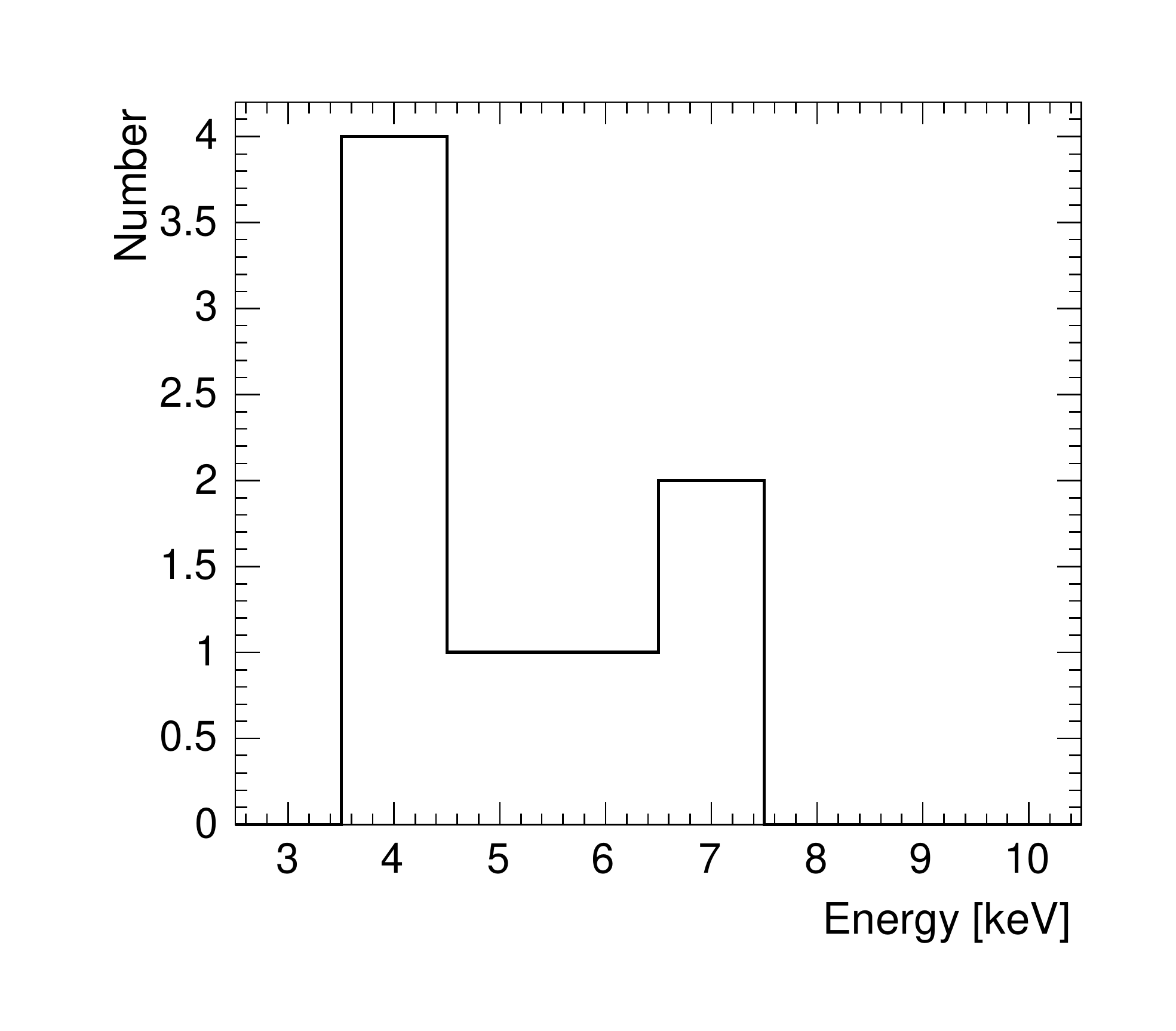}
\caption{\label{centroids} Distribution of the rest frame centroid energies 
for all lines of image~A of \rxj ~detected at a $>99\%$ confidence level. 
}
\end{figure}

Photon packets impinging on the accretion disk at $\theta=\pi/2$  are absorbed, scatter, or 
prompt the emission of an Fe~K$\alpha$ photon. We adopt a phenomenological parameterization 
for the relative probabilities of these three processes with an absorption probability of $p_{\rm abs}=0.9$ 
per encounter,  and equal probabilities for scattering and for the production of a Fe~K$\alpha$ photon packets.
Scattering off the accretion disk is implemented by first transforming the photon packet's wave and polarization 
vectors from the BL coordinates into the reference frame of the accretion disk plasma.
Subsequently, the photon packet scatters as described by the formalism of 
\citet{1960ratr.book.....C} for the reflection of polarized emission off an indefinitely 
thick electron atmosphere. After scattering, the photon packet's wave and polarization vectors 
are back-transformed into the global BL coordinate frame.  

The emission of Fe~K$\alpha$ photon packets is implemented in a similar way. 
After transforming the wave and polarization vectors of the photon packet impinging onto
the accretion disk into the reference frame of the accretion disk plasma, we use the packet's 
0-component $k^0$ of the wavevector in the plasma frame to calculate the statistical weight of 
the emitted Fe~K$\alpha$ photon packet according to the assumed power law distribution 
of the coronal emission (weight $\propto$ $(k^0)^{\Gamma-1}$). 
The mono-energetic Fe~K$\alpha$ photon packet is emitted with a limb brightening weight 
and an initial polarization given by Chandrasekhar's results for the emission of an 
indefinitely deep electron scattering atmosphere \citep{1960ratr.book.....C}. 
In the final step, the packet's wave and polarization vectors are back-transformed 
into the global BL frame.
Photons are tracked until their radial Boyer Lindquist coordinate drops below 1.02 times the $r$-coordinate 
of the event horizon (at which point we assume that the photon will disappear into the black hole) or reach
a fiducial observer at $r_{\rm obs}\,=$ 10,000~$r_{\rm g}$. In the latter case, the wave and polarization vectors are
transformed into the coordinate system of a coordinate stationary observer, and the photon-packets'
position and wavevectors are stored.

For an observer at coordinates $r_{\rm obs}$, $\theta_{\rm obs}$, and $\phi_{\rm obs}$ we select all photons
recorded in a $\theta$-window from $\theta_{\rm obs}\pm2.5^{\circ}$ and arbitrary $\phi$ 
(making use of the problem's azimuthal symmetry), and backproject each ray onto a plane 
at 10,000 $r_{\rm}$ from the observer assuming a flat spacetime to create a virtual polychromatic image.
The virtual images are subsequently convolved with the magnification maps from the RS code 
\citep[see][for a justification of the method]{Kraw:17}.
\subsection{Generation of Magnification Maps with an Ray Shooting Code}
\label{IR}

\begin{figure*}[tbh]
\begin{center}
\includegraphics[width=0.9\textwidth]{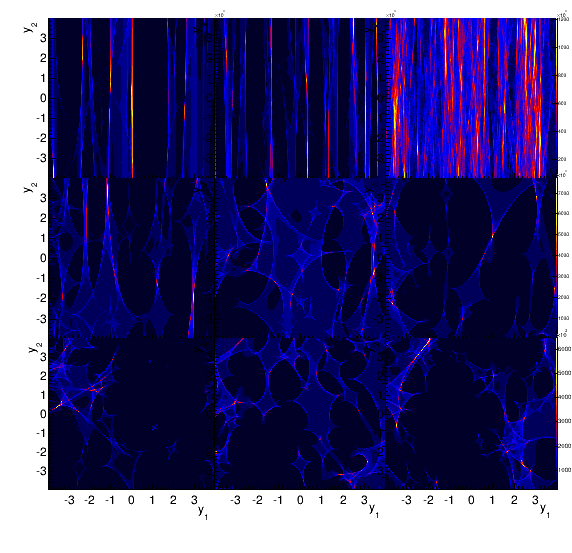}
\vspace*{-1cm}
\end{center}
\caption{\label{mm1} Simulated magnification maps for the bright saddle point (HS) image A of RX J1131$-$1231 
for different splits of the total convergence $\kappa$ between stellar convergence $\kappa_*\,=$ $g_* \kappa$
and the smooth convergence $\kappa_{\rm c}\,=$ $(1-g_*)\kappa$ with 
$g_*\,=$  0.025, 0.05, 0.1, 0.25, 0.5, 0.75, 0.9, 0.95, and 0.975 (starting in the upper left corner and going
row-wise from the left to the right).}
\end{figure*}
\begin{figure*}[tbh]
\begin{center}
\includegraphics[width=0.9\textwidth]{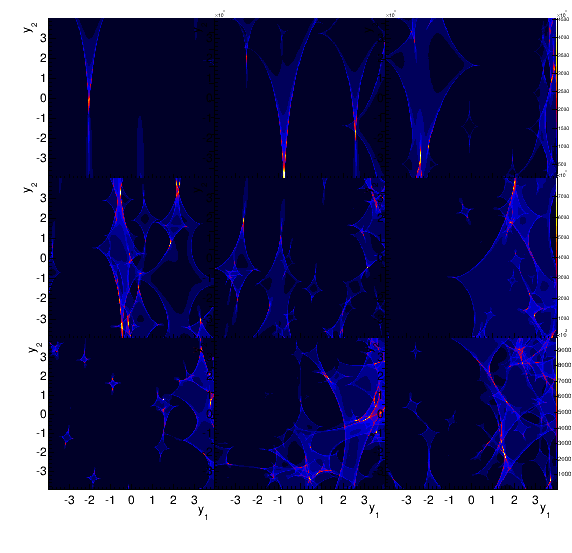}
\vspace*{-1cm}
\end{center}
\caption{\label{mm2}  Same as Figure \ref{mm1} but for the bright minimum (HM) image B of RX J1131$-$1231.
}
\end{figure*}

\begin{figure*}[tbh]
\begin{center}
\includegraphics[width=0.9\textwidth]{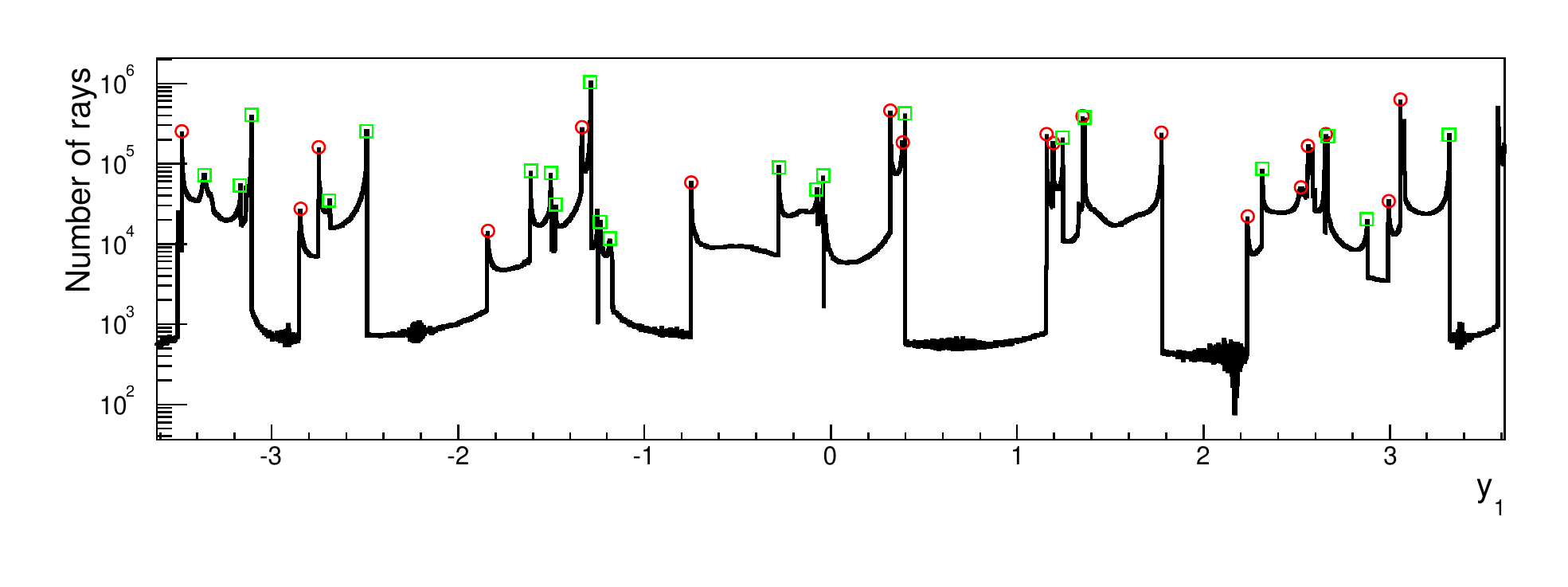}
\end{center}
\caption{\label{slice1} Slice along the $y_1$-direction (from left to right) of the magnification map of the bright
saddle point (HS) image A, showing the number of traced rays as function of $y_1$ for $g_*\,=$~5\% 
($\kappa_{\rm c}\,=$~0.42 $\gamma\,=0.597$, and $\kappa_*\,=$~0.0221). The circles and squares show 
caustics detected by a caustic finding algorithm with circles (squares) being identified as a caustic when 
searching for magnification steps from left to right (right to left).
Some caustics are found twice in which case the algorithm only stores one of the detections.
}
\end{figure*}
\begin{figure}[tbh]
\begin{center}
\plotone{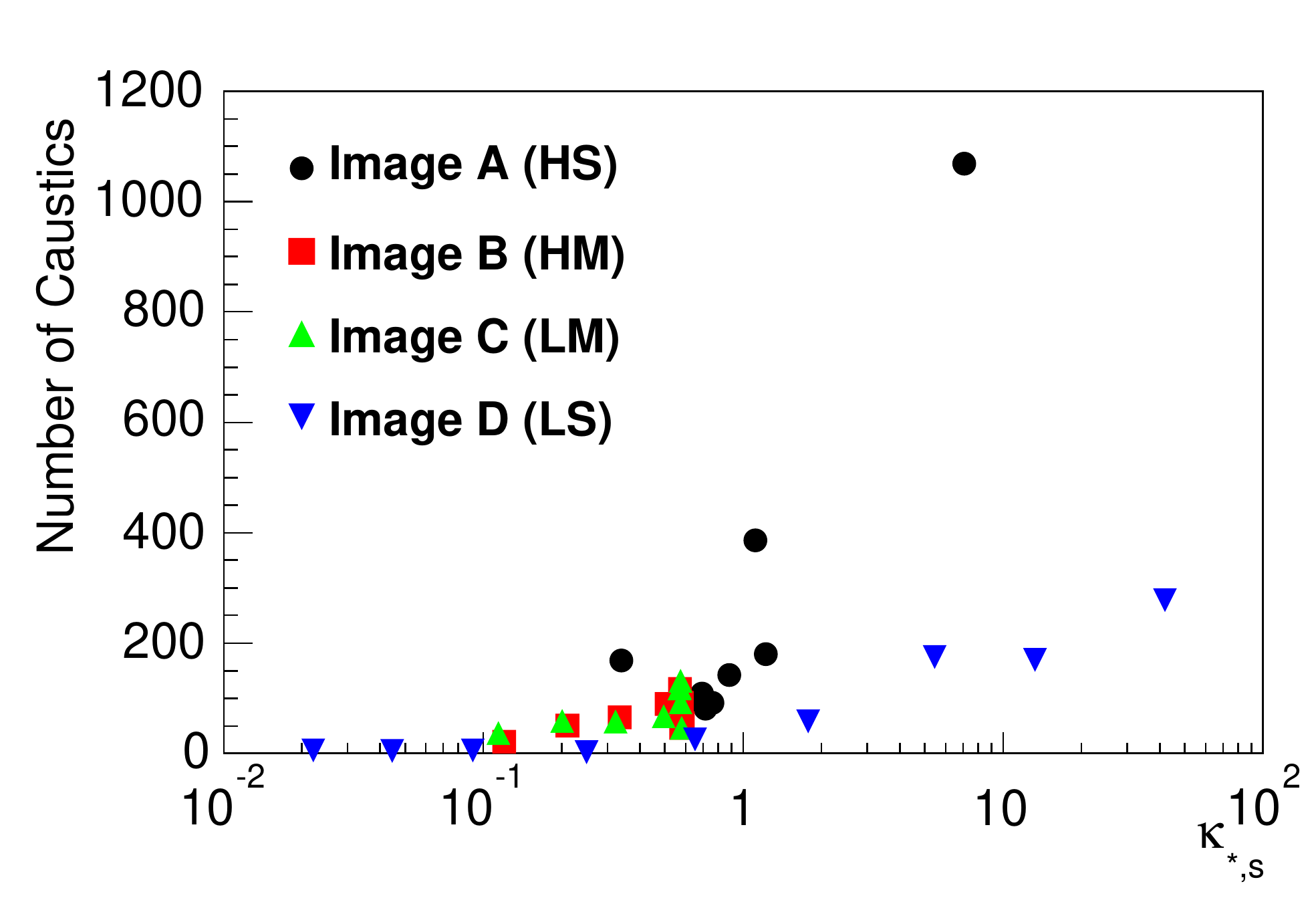}
\end{center}
\caption{\label{slice2} Number of caustics found in slices through the simulated magnification maps of images A-D of \rxj\ as function of the scaled convergence in stars $\kappa_{*,\rm s}$ (see text). The high and low saddle point images (HS, LS) behave qualitatively differently than the minimum images (HM, LM).}
\end{figure}
Our implementation of the RS technique is similar to that described by \citet{Schn:87}.
Microlensing magnification maps are generated by tracking photons backwards in time 
from a regular grid in the lens plane to the source plane. 
The local density of the endpoints of rays in the source plane is proportional 
to the cross section of the lens for deflecting rays from this location to the observer, 
and thus to the magnification $\mu$. 
Given the coordinate {\bf x} 
of a ray in the lens plane the source plane position {\bf y} is given by the lens equation:  
\begin{equation}
{\bf y}\,= 
\begin{pmatrix}
	1-\kappa_{\rm c}-\gamma & 0\\
	0 & 1-\kappa_{\rm c}+\gamma\\
\end{pmatrix}
{\bf x} -\sum_{i=1}^{N_*} \frac{m_i({\bf x}-{\bf x}_i)}{\left|{\bf x}-{\bf x}_i\right|^2}.
\label{lens}
\end{equation}
Here, {\bf x} is given in units of the Einstein radius in the lens plane: 
\begin{equation}
\xi_0\,=\,\alpha_0 \,D_{\rm L}
\end{equation}
and {\bf y} in units of the Einstein radius in the source plane:
\begin{equation}
\zeta_0\,=\,\alpha_0 \,D_{\rm S}.
\end{equation}
The first term on the right side of the lens equation models the light deflection 
by the galaxy in quadrupole approximation with $\kappa_{\rm c}$ being the surface density of 
the smooth matter distribution and distant stars in units of the critical density:
\begin{equation}
\Sigma_{\rm 0}\,=\,\left(\frac{4\pi G}{c^2}\frac{D_{\rm L}D_{\rm LS}}{D_{\rm S}}\right)^{-1}
\end{equation}
and $\gamma$ is the shear parameter. 
The second term models the deflection from 
$N_*\,=\kappa_*A_{\rm L}/\pi$ nearby stars randomly distributed over a lens plane area $A_{\rm L}$
with the masses $m_i\!\!\!<\!\!\!\!M_*\!\!\!\!>$ with $\frac{1}{N}\sum m_i=1$. 
The stellar massses $m_i$ are generated according to a power law mass function
$dN/dm \propto m^{-1.3}$ for $m\in \left[m_{\rm min},m_{\rm max}\right]$ with $m_{\rm max}/m_{\rm min}\,=\,50$, 
matching the Galactic disk mass function \citep{Goul:00,Poin:10}. 
The angular diameter distance of a source at redshift $z_2$ seen by an observer at redshift $z_1$ is given 
by \citep{Peeb:93,Hogg:00}:
\begin{equation}
D(z_1,z_2)\,=\,\frac{c}{H(z_1)}\,\frac{1+z_1}{1+z_2} \, \int_{z_1}^{z_2} \frac{dz'}{E(z')}
\end{equation}
with $E(z)$ and $H(z)$ given by:
\begin{equation}
E(z)\,=\,\,\sqrt{\Omega_{\rm m}(1+z)^3 +\Omega_{\rm \Lambda}}.
\end{equation}
and
\begin{equation}
H(z)\,=\,H_0 \,E(z).
\end{equation}
For \rxj\ and $<\!\!\!M_*\!\!\!>$~$=\,0.1\, M_{\odot}$ we obtain:
$D_{\rm L}=$2.9$\times 10^{27}$~cm,
$D_{\rm S}=$4.6$\times 10^{27}$~cm,
$D_{\rm LS}=$2.6$\times 10^{27}$~cm,
$\xi_0=$9.8$\times 10^{15}$~cm, and
$\zeta_0=$1.5$\times 10^{16}$~cm.
Assuming a black hole mass of $10^8\,M_{\odot}$, 
$\zeta_0=1046\,r_{\rm g}$ ($r_{\rm g}=0.000956\, \zeta_0$). 

The rays are generated over a large lens plane ``source area'' of (dimensionless) width $L_1$ and height $L_2$ to cover an approximately square-shaped ``target area'' of (dimensionless) width and height $L_{\rm s}\,=8$ in the source plane. The source area is chosen to be sufficiently large so that 
rays originating outside of the source area have a negligibly small likelihood of ending 
in the target area. For this purpose, we start with 
$L_1\,\approx$ $f_0\, f_1\,L_{\rm s}$ and $L_2\,\approx$ $f_0\,f_2\,L_{\rm s}$.
The factor $f_0\,=$5 assures that the rays cover a sufficiently large area in the source plane 
(much larger than actually needed), and the factors
$f_1\,=\, \left| 1-(\kappa_{\rm c}-\gamma)\right|^{-1}$ and 
$f_2\,=\, \left| 1-(\kappa_{\rm c}+\gamma)\right|^{-1}$ account 
for the scaling of distances between the source and lens planes.
We distribute stars over a 4 times larger lens plane area to make sure that all rays are 
surrounded by a large number of stars.
Two methods are used to speed up the calculation \citep[see][for similar approaches]{Schn:87,Wamb:99}. 
In the first iteration, we shoot 1 million rays and tag all the lens plane locations whose rays 
end up in the target area.  The second iteration is then limited to regions surrounding 
the tagged portions of the lens plane. The second ray shooting iteration uses a finer mesh, 
i.e.\ we shoot 200$\times$200 rays for each tagged ray of the first iteration.  
The shooting of the 200$\times$200 rays spread over an area $(\Delta x)^2$ 
in the lens plane ($\Delta x$ being the lens plane distance between 
adjacent rays of the first iteration) is accelerated by dividing the stars into nearby stars 
(distance $<30\,\Delta x$) and distant (all other) stars. 
For each ray the deflections from all nearby stars are calculated exactly, and the deflections 
from all other stars and the macrolens are calculated with a bilinear interpolation using the 
deflections at the four corners of the considered area.
For all our calculations, we check the accuracy of the interpolation scheme 
for a small fraction of all rays by calculating the deflection by straight summation over all stars. 
We find the errors to be negligibly small ($<0.01$ in units of the gravitational radius of the black hole).
We store ``overview'' magnification maps on different scales 
(side lengths of 100, 10, 1, and 0.02 in dimensionless units), 
sample slices through the maps, and forty 400\,$r_{\rm g}$ diameter 
regions randomly distributed over the central portion of the magnification maps.
Random portions of the latter maps are subsequently folded with the virtual quasar images
to generate microlensed energy spectra.
\section{Properties of the Ray Shooting Magnification Maps}\label{maps}
We generated magnification maps using for each image the total convergence $\kappa$ and shear $\gamma$  
from the analysis of \citep{Black:11,Pool:12}.
The authors derived the lens parameters with the code of \citet{Keet:01} based on the positions of the 
macroimages and neglecting the observed fluxes and time delays. 
Using the same approach as \citet{Pool:12}, we generated for each image a series of magnification maps 
using different combinations of the smooth convergence $\kappa_{\rm c}=(1-g_*)\kappa$ 
and the convergence from the stellar component  $\kappa_*\,=\,g_* \kappa$ with 
$g_*\,=$ 0.025, 0.05, 0.1, 0.25, 0.5, 0.75, 0.9, 0.95, and 0.975.

Figure \ref{mm1} shows the magnification maps for the bright saddle point (HS) image A 
($\kappa\,=$~0.442, $\gamma\,=$~0.597). 
The density of the caustics in the source plane increases as the fraction of the convergence in stars 
increases from $g_*\,=$~2.5\% to 5\% and 10\%. For larger $g_*$-values the caustic density decreases markedly.
The behavior contrasts with that of Figure 2 of \citet{Sche:02} where the caustic density increases 
monotonically with $g_*$.  
The difference is that the macro-models considered here have a substantial shear and 
very large magnification factors $\left|1/(1-(\kappa+\gamma))\right|\,\approx\,$26 along one direction.
We find that for constant overall convergence and shear, the model with the largest 
stellar density per dimensionless source plane area (accounting for the scaling 
between distances in the lens plane and in the source plane):
\begin{equation}
\kappa_{*,\rm s}\,=\,\frac{\kappa_*}{\left|\left(1-\kappa_{\rm c}-\gamma\right) \left(1-\kappa_{\rm c}+\gamma\right)\right|}
\label{kappaSourcePlane}
\end{equation}
produces the highest caustic density in the source plane. 
We obtain a better correlation of $\kappa_{*,\rm s}$ with the 
caustic density when scaling between the lens plane and source plane areas 
with $\kappa_{\rm c}$ rather than with the overall convergence $\kappa=\kappa_{\rm c}+\kappa_*$ 
in the denominator of Equation (\ref{kappaSourcePlane}).
The stellar convergence $\kappa_*$ contributes to focusing light rays, but does
not contribute in the same way to the scaling between the density of stars 
and the density of caustics in the source plane. 
The source plane statistics of caustics for the cases of $\kappa\pm\gamma\approx1$
seems to be an interesting topic for future investigations.

Figure \ref{mm2} shows the magnification maps for image B ($\kappa\,=$~0.423, $\gamma\,=$~0.507).
In accord with the earlier results of \citet{Witt:95,Metc:01,Sche:02}, we find that the magnification values 
vary more markedly for the saddle point image A (Figure \ref{mm1}) than for the minimum image B 
(Figure \ref{mm2}).
We developed an algorithm to identify caustics in 30 $r_{\rm g}$ wide slices through the 
caustics maps running horizontally from left to right in Figs.\ \ref{mm1} and \ref{mm2}.
The algorithm exponentially averages the magnification running from 
left to right (averaging over all bins to the left of a considered point, using for each point a weight 
which decreases exponentially with the distance from the point) and from right to left, 
and then identifies caustics through the peaks in the difference distribution.  
Figure \ref{slice1} shows the magnifications for an exemplary slice of the magnification 
map of Fig.\ \ref{mm1} together with the caustics that the algorithm identified. 
Figure \ref{slice2} summarizes the results from analyzing slices through all the magnification maps, namely
the number of detected caustics as a function of $\kappa_{*,\rm s}$. 
We see that the number of caustics indeed correlate well with $\kappa_{*,\rm s}$.
Each image shows a slightly different correlation, with those of the saddle point 
images A and D exhibiting a qualitatively different behavior than the 
minimum images B and C. The scatter in the distribution comes from the finite 
size of the simulated magnification maps.
\begin{figure*}[tbh]
\begin{center}
\includegraphics[width=0.9\textwidth]{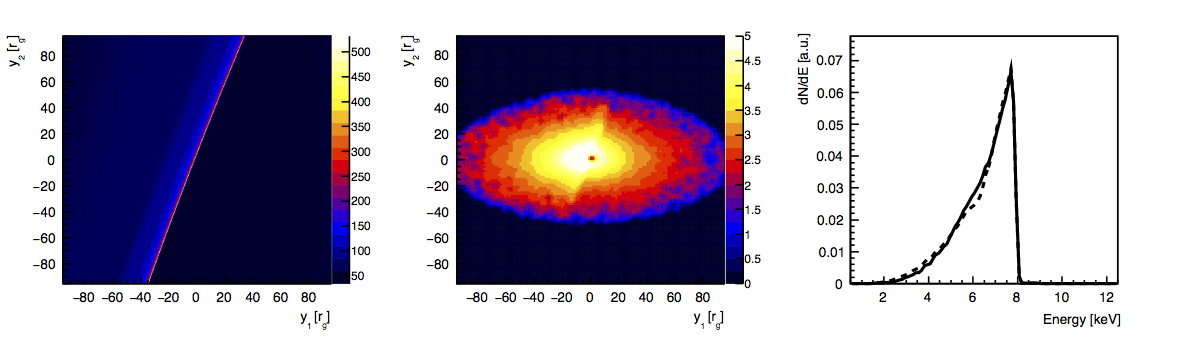}
\includegraphics[width=0.9\textwidth]{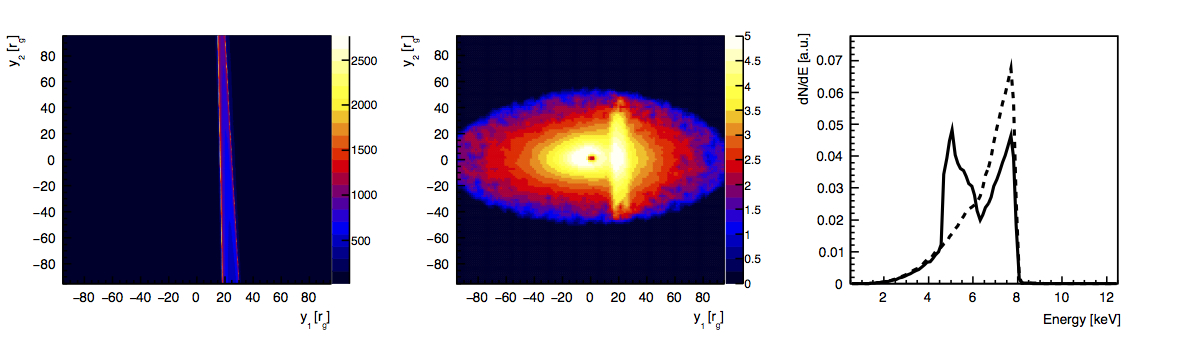}
\includegraphics[width=0.9\textwidth]{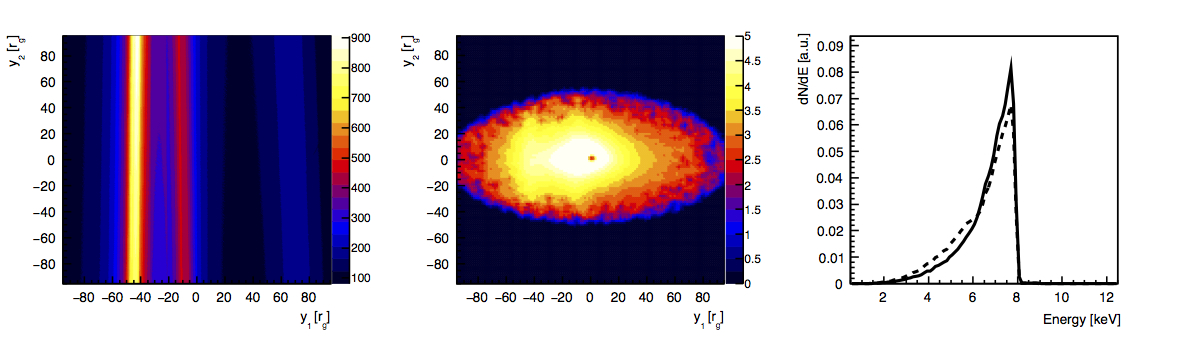}

\end{center}
\caption{\label{comb1} The left panels show small regions of the magnification map for image A of \rxj, 
with $g_*\,=$~5\% ($\kappa_{\rm c}\,=$~0.42 $\gamma\,=0.597$, and $\kappa_*\,=$~0.0221).
The center panels show the virtual images of the Fe~K$\alpha$ surface brightness convolved with the
magnification pattern from the left panels. 
The right panels show the resulting Fe~K$\alpha$ energy spectra
before (dashed lines) and after (solid lines) accounting for the microlensing magnification.
We normalize all energy spectra to the same total flux to facilitate the comparison 
of their shapes. All figures are shown for spin parameter of $a=0.9$, 
a lamppost corona at $h = 5r_g$, and an inclination angle of $i = 62.5^{\circ}$.  
 }
\end{figure*}
\section{Simulation of the Microlensed Fe~K$\alpha$ Emission\label{fka}}
In the previous section we showed the magnification maps for rather large source plane regions, 
when comparing the width or height ($L_s\,=\,8$ in dimensionless source plane units) of the maps 
with the size scale of the inner accretion disk of several $r_{\rm g}$ (with $r_{\rm g}\approx$~0.001 
for \rxj\ in dimensionless source plane units). 
Simulating such large maps is not a choice but a necessity for assuring that the RS method uses a
sufficiently large number of stars to lead to acceptable small number statistics and edge effects.
In this section we present magnification maps on the relevant spatial scales of a few 10 $r_{\rm g}$'s, 
and explore how they distort the energy spectra of the microlensed Fe~K$\alpha$ emission
for a black hole with $r_{\rm g}=$~0.00096 in dimensionless units, spin $a=0.9$ in geometrical units, 
a lamppost corona at $h=5r_{\rm g}$.
Furthermore, we assume that the shear direction is aligned with the spin axis of the black hole, as 
caustics perpendicular to the accretion disk tend to maximize the spectral distortions.

\begin{figure}[t]
\plotone{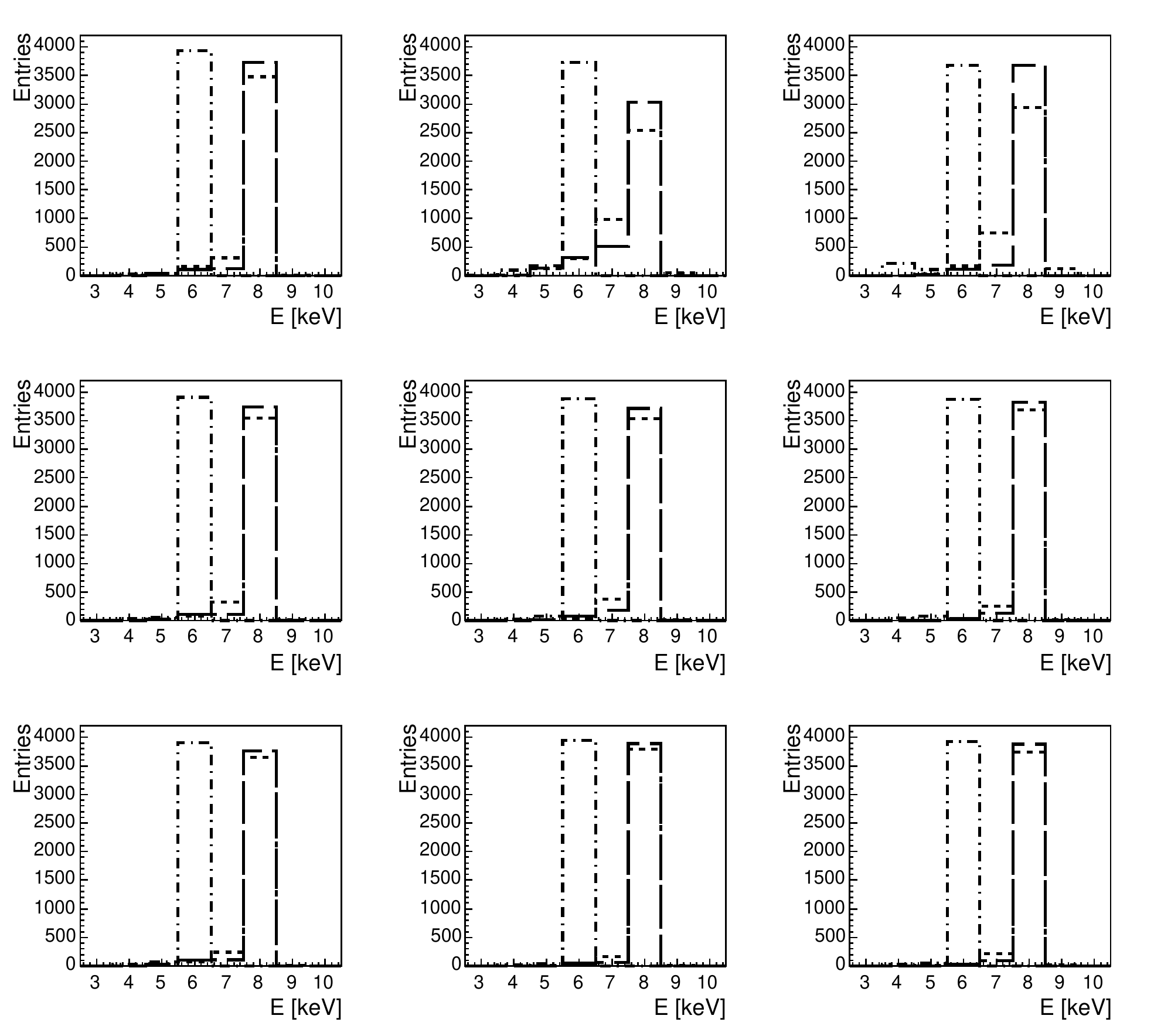}
\caption{\label{dist1} Distribution of the line centroid energies for the simulations of 
image A of \rxj\ for the microlensing models shown in Figure \ref{mm1} and a black hole inclination of 
2.5 $^{\circ}$ (dot-dashed line), 
62.5$^{\circ}$ (long-dashed line) and
82.5$^{\circ}$ (short-dashed line) for different splits of the total convergence $\kappa$ 
between stellar convergence $\kappa_*\,=$ $g_* \kappa$ and smooth convergence 
$\kappa_{\rm c}\,=$ $(1-g_*)\kappa$ with 
$g_*\,=$  0.025, 0.05, 0.1, 0.25, 0.5, 0.75, 0.9, 0.95, and 0.975 
(starting in the upper left corner and going
row-wise from the left to the right).
As in the previous figures, a spin parameter of $a=0.9$ is chosen 
and the lamppost corona is located at $h = 5r_g$.  
The energies are give in the quasar reference frame.
}
\end{figure}

Figure \ref{comb1} shows random locations of the magnification maps (left panels), 
the Fe~K$\alpha$ emissivity convolved with these magnification maps (center panels), 
and the resulting Fe~K$\alpha$ energy spectra (right panels) for Image A with $g_*\,=$~5\%.
The upper panel of Figure \ref{comb1} shows a map with a rather simple caustic structure 
selectively amplifying the emission from certain portions of the accretion disk leading to a somewhat 
distorted energy spectrum. The center panel of Figure \ref{comb1} shows that the microlensing can produce
energy spectra with multiple pronounced peaks.
The lower panel of Figure \ref{comb1} shows that many caustics do not necessarily 
lead to more extreme spectral distortions, as the superposition of the caustics leads 
to a more uniform magnification of the disk emission than a single caustic.

We perform a qualitative analysis of the simulated Fe~K$\alpha$ energy spectra with a simple algorithm 
finding the highest peak of each simulated energy spectrum, and searching for a secondary peak 
with a peak flux value exceeding by 4\% of the primary peak flux amplitude 
in the valley between the primary and secondary peaks.
Figure \ref{dist1} shows the distribution of the peaks of the Fe~K$\alpha$ energy spectra as function of $g_*$
(image A of \rxj\ and black hole inclinations of 2.5$^{\circ}$, 62.5$^{\circ}$ and 82.5$^{\circ}$). The three energy 
spectra with $g_*\le$ 10\% are most strongly affected by the micronlensing. 
The models for $g_*\,=$~5\% and $g_*\,=$~10\% show that the microlensing has some - albeit limited -
impact on the line centroid distribution.  

\begin{figure}[tb]
\plotone{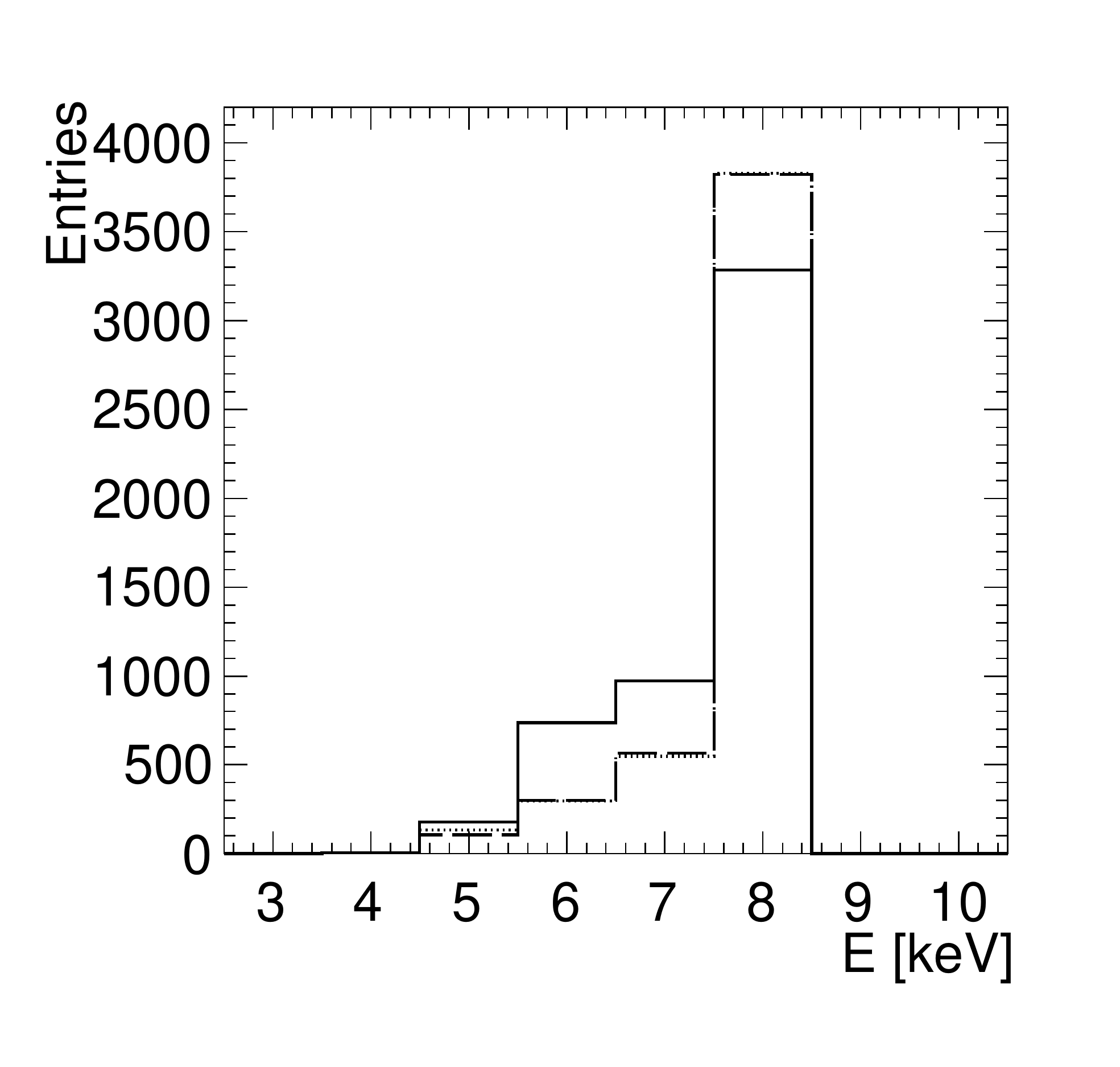}
\caption{\label{dist6} Distribution of the line centroid energies for the simulations of 
image A of \rxj\ with $g_*\,=$~5\% for a black hole spin $a$ of 0 (solid line), 0.9 (long-dashed line), and 0.998 (dotted line).
All distributions assume a lamppost corona at $h = 5r_g$ and an inclination angle of $i = 62.5^{\circ}$.  
The energies are give in the quasar reference frame.}
\end{figure}
\begin{figure}[tb]
\vspace*{0.5cm}
\plotone{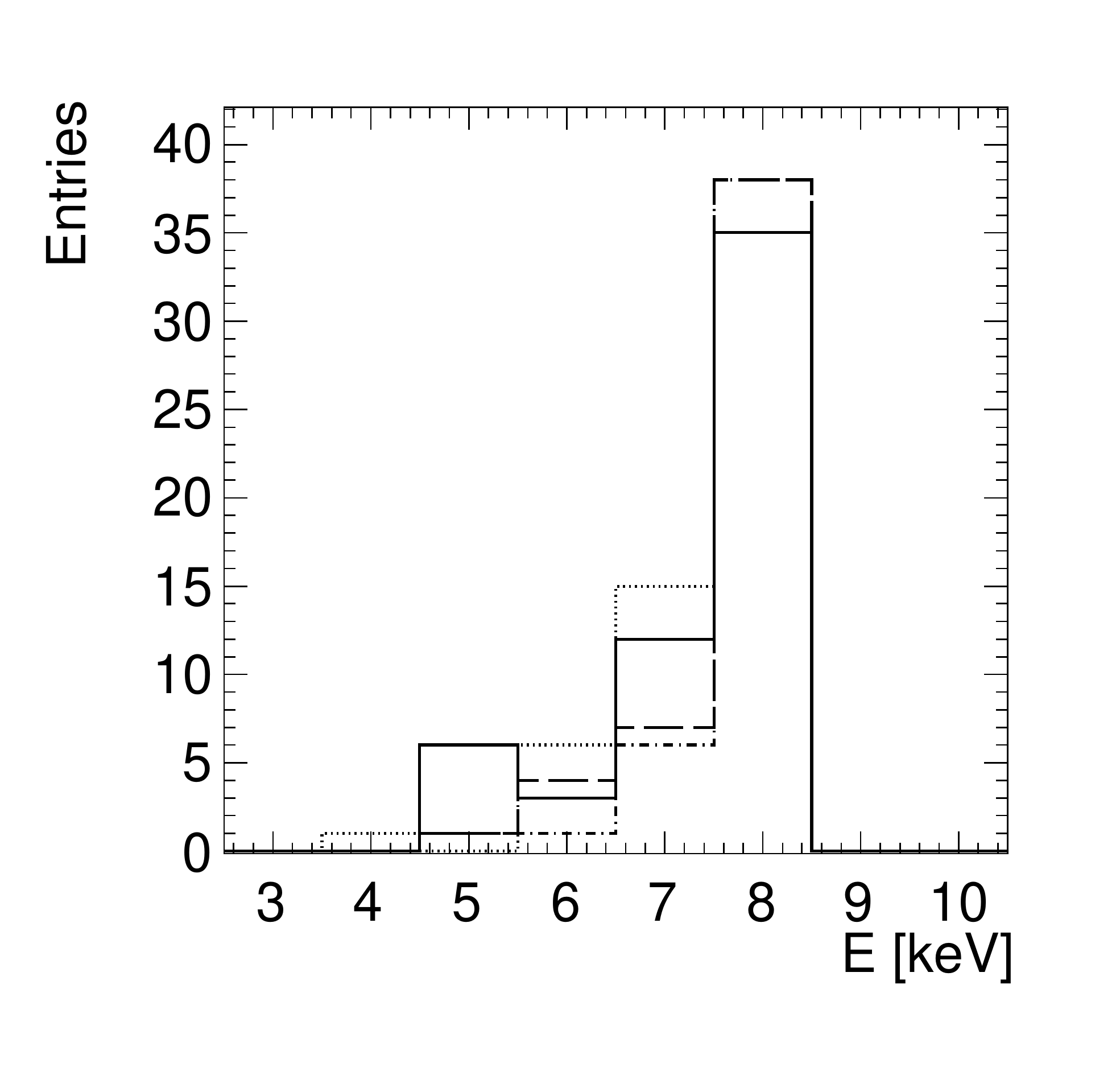}
\caption{\label{dist5} Distribution of the line centroid energies for the simulations of 
image A of \rxj\ with $g_*\,=$~5\% scaling the mass of the black hole by $\lambda$, or, alternatively with mass scale 
of the deflecting stars by $\lambda^{-1/2}$ with $\lambda\,=$~4 (solid line), 2 (dotted), 1 (long-dashed),  and 0.67 (dashed-dotted).  
All distributions assume a black hole spin parameter of $a=0.9$, a lamppost corona at $h = 5r_g$,
and an inclination angle of $i = 62.5^{\circ}$. The energies are give in the quasar reference frame.  
}
\vspace*{0.5cm}
\end{figure}

Figure \ref{dist6} presents the distribution of the peaks of the Fe K$\alpha$ emission 
for several black hole spins.
The distribution is always strongly peaked around the peak energy of the unlensed emission, which shifts
slightly from lower to higher energies as the black hole spin increases.
Figure \ref{dist5} shows how the distribution changes when the mass of the black hole 
(giving physical size of the accretion system) is changed by a factor $\lambda$, 
or, alternatively the mass scale of the deflecting stars is changed by a factor $\lambda^{\-1/2}$.
Although $\lambda$ does impact the distribution somewhat, the main characteristic (i.e.\ a pronounced peak at the energy
of the peak of the distribution without gravitational lensing) remains unaffected.
 
As expected from Figure \ref{mm2}, the minimum image B (and similarly image C) is much less affected 
by microlensing as the source plane density of caustics is much smaller, and it is correspondingly less likely that the 
central portion of the accretion disk intercepts a caustic fold or cusp. 
The peak energies (not shown here) are narrowly distributed around  the value without gravitational lensing.
\section{Comparison Between Simulations and {\it Chandra} Observations\label{chandra}}
The simulated line centroid distributions of Figures \ref{dist1}-\ref{dist5} 
cannot be directly compared to the observed distribution of Fig.~\ref{centroids}, 
as the latter are affected by detection biases resulting for example 
from {\it Chandra's} effective area and energy resolution, the energy dependent
signal to noise ratio and the particular analysis choices.
We performed a first analysis that accounts for these biases by converting
the simulated Fe~K$\alpha$ energy  spectra into simulated {\it Chandra} data sets,
and by applying an automized search for emission lines to these data sets.

\begin{figure}[t]
\plotone{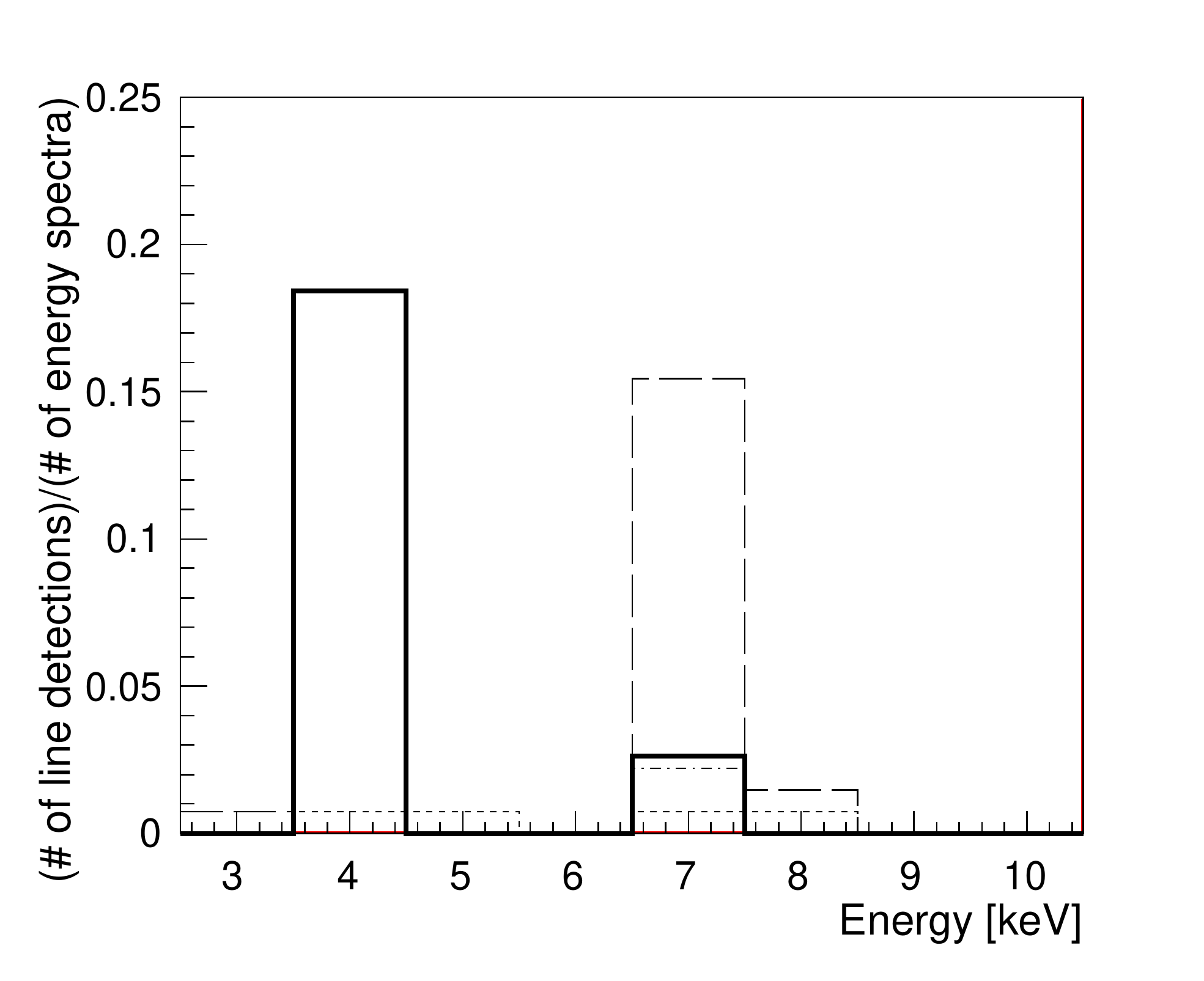}\vspace*{-0.25cm}
\plotone{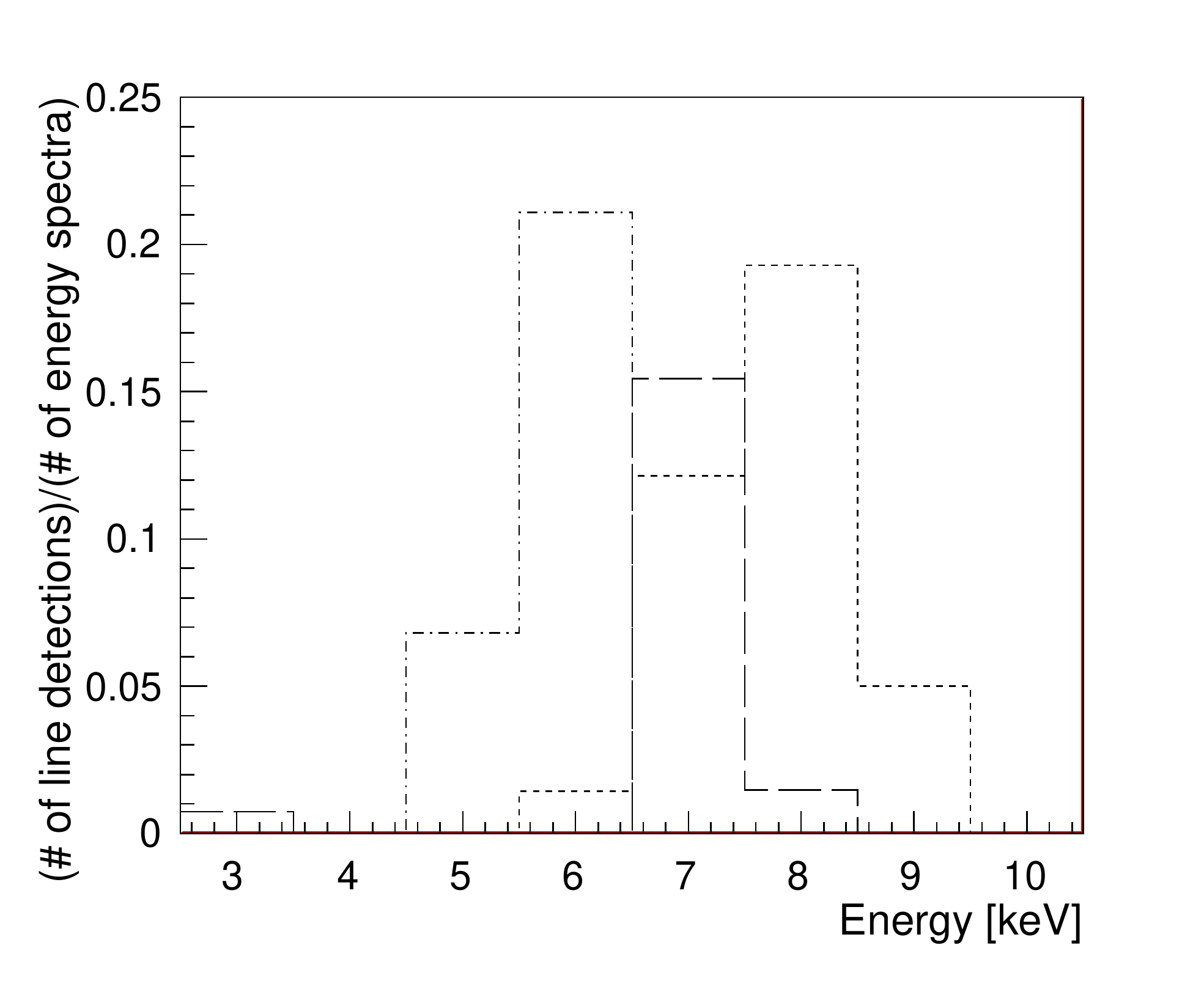}\vspace*{-0.25cm}
\plotone{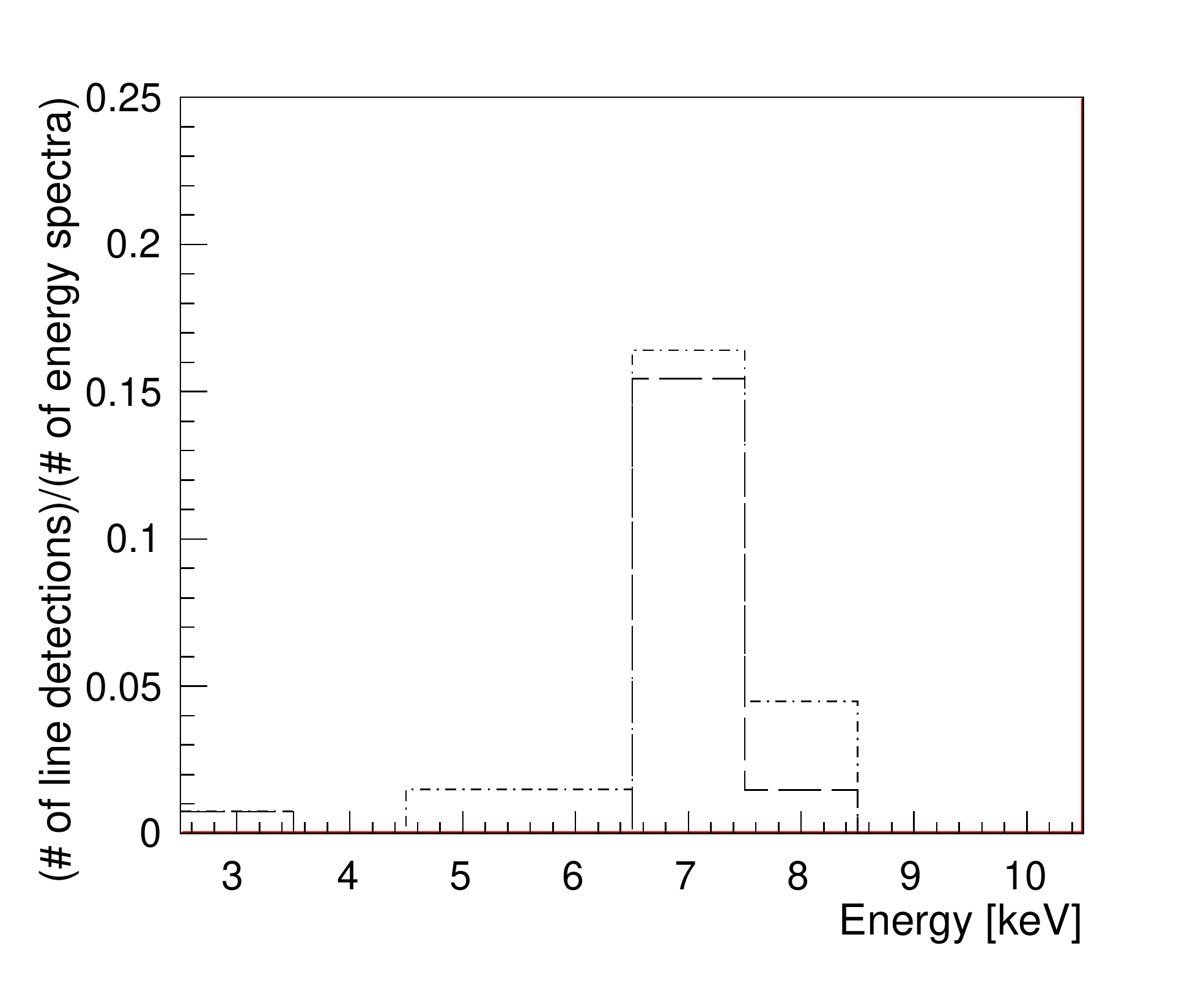}\vspace*{-0.2cm}
\caption{\label{c1} The upper panel shows the line energies found by the automated analysis 
script in the observed \rxj\ data set (thick solid line) and in the simulated data sets (thin lines) 
for image A of the source (all energies are in the quasar reference frame).  
The simulations are for a black hole spin of $a=0.9$, inclination of $i=62.5^{\circ}$, 
and microlensing with $g_*=5\%$.
The thin lines show the results for different choices of the line intensities 
varying by a factor of two between adjacent lines.
The center panel presents simulated results for black hole inclinations of 
2.5$^{\circ}$ (dashed-dotted line),
62.5$^{\circ}$ (long-dashed line), and
82.5$^{\circ}$ (short-dashed line).
The lower panel shows simulated results for $i=62.5^{\circ}$ for two different 
microlensing parameters: $g_*=$ 0.025 (dashed-dotted line) and 0.05 (long-dashed line).
}
\vspace*{-0.2cm}
\end{figure}

The HEASARC (High Energy Astrophysics Science Archive Research Center) software 
(i.e.\ the {\it table} class of the HEASP package) is used to generate a table model 
for each simulated Fe K$\alpha$ energy spectrum. A model consisting of an absorbed
powerlaw model plus the Fe K$\alpha$ energy spectrum is subsequently defined   
in the {\it Xspec} fitting package \cite{Arna:17}. The {\it Xspec} command {\it fake} is 
subsequently used to generate a {\it Chandra} energy spectrum. 
Hereby the overall normalization of the Fe K$\alpha$ line flux is treated as an 
adjustable parameter. 

The automatic search for Fe~K$\alpha$ lines is performed with a 
Tool Command Language ({\it TCL}) script. The script first fits an absorbed powerlaw model. 
Subsequently, it iterates over the starting value $E_1$ of the centroid energy of an added 
Gaussian emission line from 1.5 keV to 6 keV in 0.1 keV steps (observer frame energies).
The script identifies the best fit and estimates the statistical evidence for a line detection  
from Fisher statistic. The iteration over $E_1$ is followed by the iteration over the line centroid energy 
$E_2$ of a second line.  The algorithm again identifies the best fit and evaluates Fisher statistic 
to obtain a measure for the statistical significance of the detection of a second line.

It is well known that the Fisher statistic under or overestimate the chance probability of 
certain spectral features \cite{Prot:02}. This does not matter here, as we are only 
interested in the comparison of the results obtained for the observed and simulated data. 
We choose a threshold value of the Fisher statistic, so that the automatic script gives 
approximately the same number of lines when applied to the observed {\it Chandra} data 
as the manual analysis.

The automated analysis is identically applied to the observed and simulated data sets.
The thick solid line in the upper panel of Figure \ref{c1} shows the distribution of the 
lines detected by the automated script in the observed spectra of \rxj\.
The results can be compared to those from the manual analysis in Fig.\ \ref{centroids}.
Although both analyses find most line centroids at source frame energies around $\sim$4 keV, 
the automated script finds a significantly larger fraction of lines at 4 keV than at 7 keV in 
image A of \rxj\ compared to the manual spectral analysis. 
Part of the difference results from the fact that the confidence levels are derived more 
rigorously in the manual approach which makes use of Monte Carlo simulations 
than the automated approach which use the less reliable Fisher statistic.

From bottom to top, the dotted (barely visible), short-dashed, dashed-dotted, and long-dashed 
lines show the spectral lines found in the simulations for image A for different Fe K$\alpha$ 
emission strengths increasing from line to line by a factor of 2.  The data and simulation histograms 
are normalized to the number of analyzed energy spectra, so that the shapes and absolute 
values of the distributions  can be compared to each other. If we adjust the flux normalization 
of the Fe K$\alpha$ emission so that we get approximately the same number of detections 
as in the observed \rxj\ data set, the simulations predict line detections narrowly clustered 
around line energies of 7 keV. The distribution deviates significantly from the distribution of 
lines detected by the manual and automated analyses in the real {\it Chandra} data. 
In particular, the simulations do not reproduce the frequent detection of lines around 4 keV.

The center panel of Fig.~\ref{c1} shows the fitted line centroids for different simulated
inclinations. For each inclination we chose a representative Fe~K$\alpha$ intensity which gives
roughly the right number of line detections. At the lower inclinations, the line centroids 
shift towards lower energies, but still do not reproduce the large number of lines detected around 4 keV.

The lower panel of Fig.~\ref{c1}  shows that the simulations for different $g_*$-values 
give very similar line centroid distributions -- once we adjust the Fe~K$\alpha$ intensity appropriately. 
The microlensing has a rather limited impact on the detected line centroid distribution.

\begin{figure}[h]
\plotone{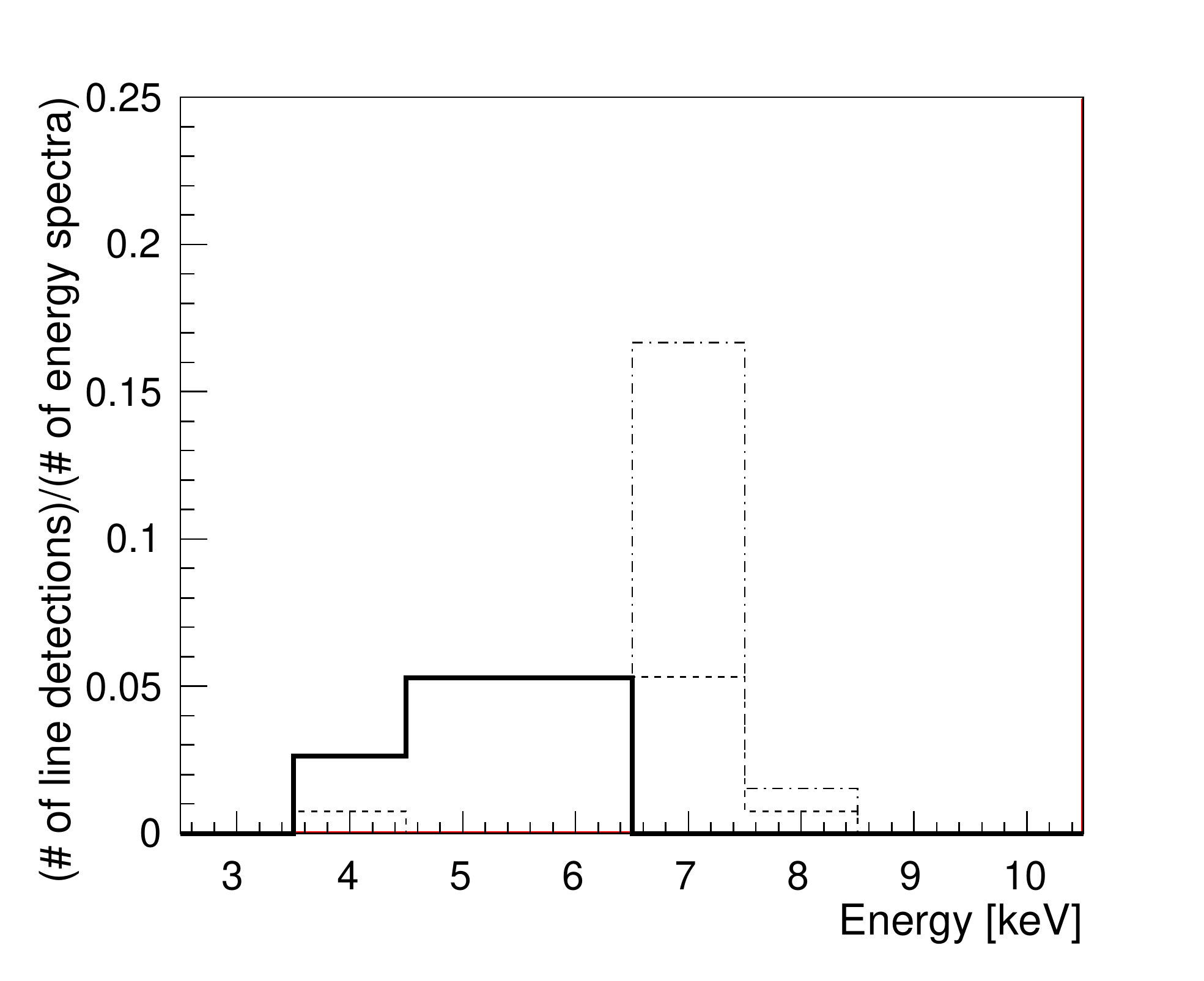}
\vspace*{0.25ex}
\plotone{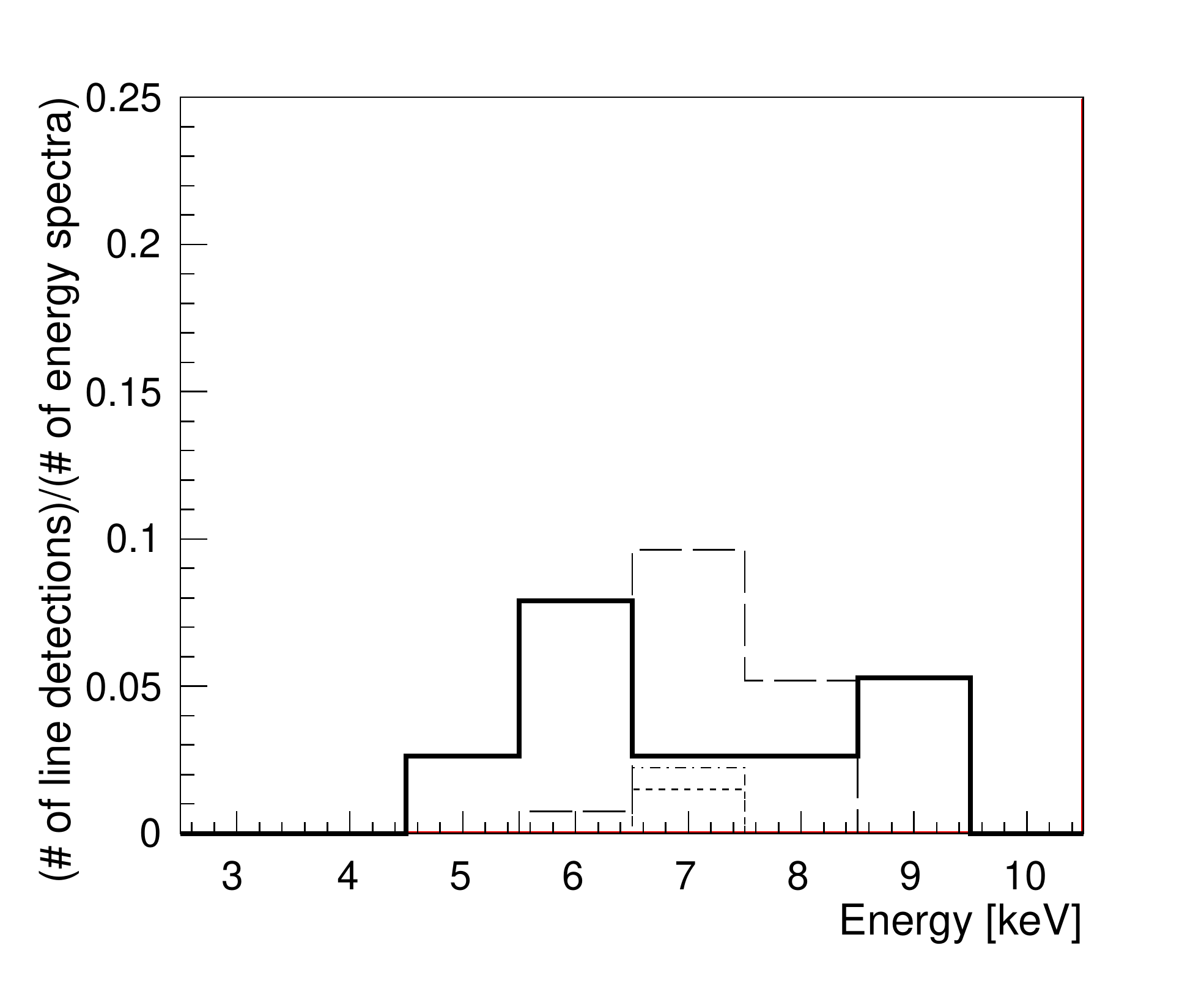}
\vspace*{0.25ex}
\plotone{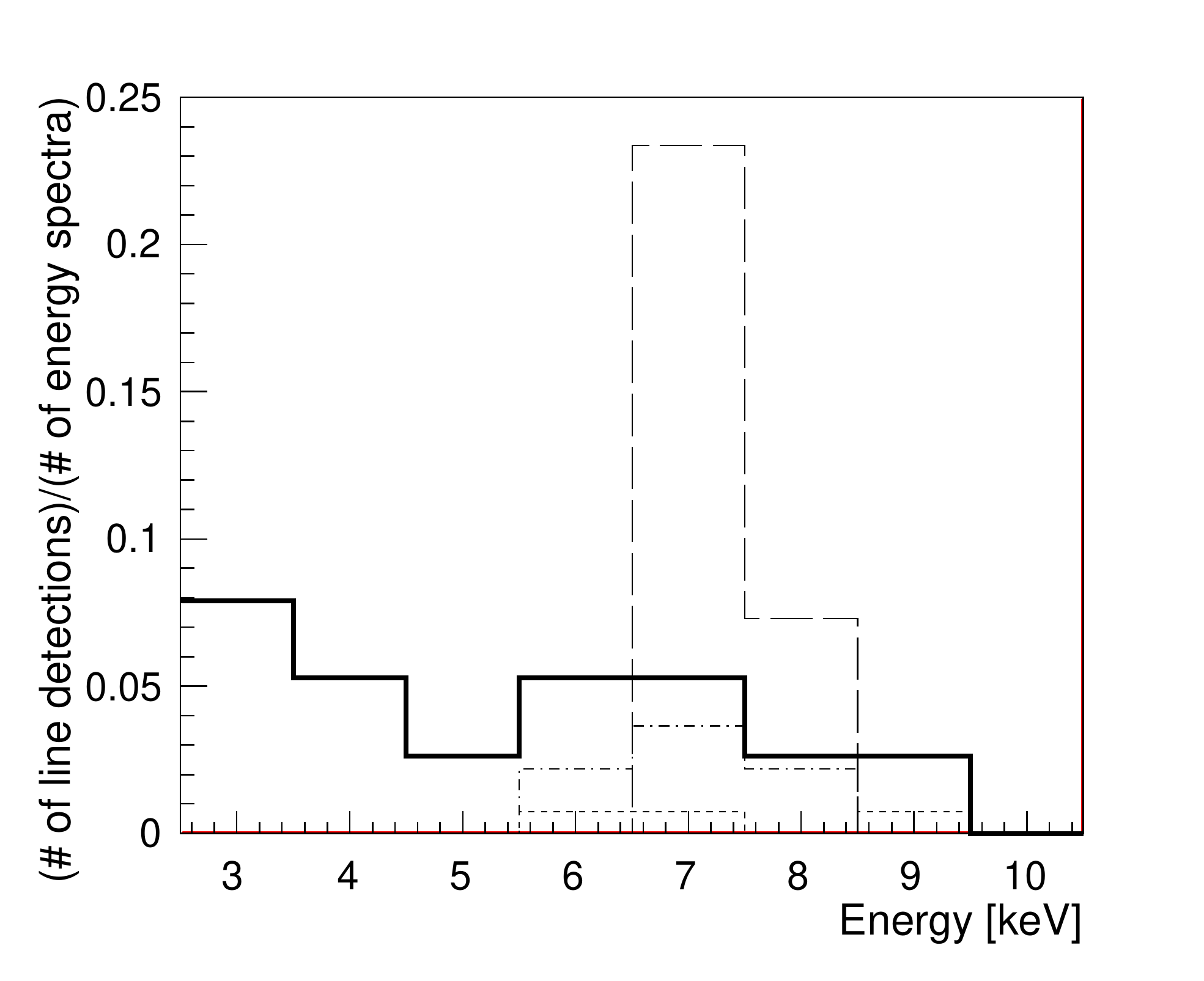}
\vspace*{0.25ex}
\caption{\label{c2} Same as the top panel of Fig.\ \ref{c1} but for images B (top), C (center) and D (bottom).}
\vspace*{0.25ex}
\end{figure}

We present the comparison between the simulated and observed line centroid distributions
for images B, C, and D in Figure \ref{c2}. For images B and D the simulated 
distributions clearly underpredict the number of line detections at $<$5~keV energies.

We checked the results for different black hole spins and got similar results. 
For lower (higher) spin parameters, the centroid distributions are slightly narrower (wider) 
than for the spin parameter $a=0.9$ - but without substantially affecting the mismatch 
between the simulations and the observations.   
\\[4ex]
\section{Summary and Discussion \label{discussion}}
It is instructive to point out similarities and differences between this paper discussing the 
distribution of the spectral shapes of the Fe~K$\alpha$ emission from microlensed quasars
and earlier papers discussing the statistical distribution of brightness fluctuations caused 
by microlensing \citep[e.g.][]{Sche:02,Pool:12} and the detailed modeling of the micrlolensed
light curves \citep[e.g.][]{Lewi:96,Dai:10}. The distribution of the magnification factors does 
not depend on the spatial (or angular) scale of the magnification maps relative to the spatial 
scale of the quasar accretion disk, as long as the Einstein radius of the deflectors is much larger 
than the sizes of the emitting regions. The unitless lens equation used to model the distribution 
of magnification factors does indeed not depend on the mass scale of the deflectors $<\!\!m_*\!\!>$. 
In contrast, the modeling of the microlensed light curves does depend on the spatial scale 
of the magnification maps, as the time between caustic crossings depends on this scale 
divided by the velocity of the deflectors perpendicular to the line of sight. 
In the case considered here, the results depend on the spatial scale of the magnification 
maps as well, as the spatial scale impacts the expected number of caustics intersecting the
central portion of the accretion flow.

The simulations show that microlensing can modify the energy spectra of the 
observed Fe K$\alpha$ emission, but the effect is not large enough to explain the 
range of the observed line energies (Figures \ref{c1} and \ref{c2}). 
In particular, the simulations do not reproduce the highly redshifted peaks with $<$5~keV 
centroids seen in the data. The overall small impact of the microlensing on the observed 
distribution of line energies has several explanations. For microlensing maps with a low
density of caustics, the caustics do not intersect the inner accretion flow often enough. 
For microlensing maps with a high densities of caustics, we observe rather small 
distortions of the Fe K$\alpha$ lines when compared to the statistical uncertainties in 
the {\it Chandra} data. Another contributing effect is that microlensing most likely leads 
to a line detection when the brightest emission is amplified, giving a line detection 
with a centroid similar to that of the unlensed Fe K$\alpha$ emission.   
   
Our simulations have several shortcomings. We have not modeled the spatial 
extent of the corona, and we did not model the microlensing of the direct corona emission. 
Furthermore, we describe the emitted Fe K$\alpha$ energy spectrum in the rest frame 
of the accreting plasma with a delta-function in energy (see \cite{Garc:16}, and references 
therein for more realistic emission energy spectra). We choose for each observation a random 
patch from a large magnification map, assuming that the caustics move relatively fast relative 
to the quasar so that each {\it Chandra} observation catches a different region of the caustic net. 
In reality, the caustics may move relatively slowly so that multiple {\it Chandra} observations 
are affected by a single caustic structure. The geometry of the accretion disk may differ from 
the paper-thin geometry assumed here. The disk may be geometrically thick, 
it may be patchy, or partially obscured. The geometry of the inner accretion flow of powerful 
sources like \rxj\ may differ from that of the well studied nearby Seyfert 1 galaxies. 
If the black hole is sufficiently more massive than what we assume, it would intersect 
a larger number of caustics. Last but not least, our microlensing model may be insufficient. 
For example, \citet{Dai:18} propose that planets in the lensing galaxy may produce caustics 
with observable signatures. We plan to evaluate the impact of these effects in our future work.

\section*{Acknowledgments}
HK would like to thank NASA (grant \#NNX14AD19G) and the 
Washington University McDonnell Center for the Space Sciences for financial support.
GC would like to acknowledge financial support from NASA via the Smithsonian Institution 
grants SAO GO4-15112X, GO3-14110A/B/C, GO2-13132C, GO1-12139C, and GO0-11121C.
HK thanks Quin Abarr for implementing the Cash-Karp integration of the geodesics into the
ray tracing code.

\end{document}